\documentclass[%
 reprint,
bibnotes,
 amsmath,amssymb,
 aps,
]{revtex4-2}

\usepackage{graphicx}
\usepackage{dcolumn}
\usepackage{bm}
\usepackage{hyperref}
\usepackage{xcolor}
\usepackage[mathlines]{lineno}

\begin{document}

\preprint{APS/123-QED}

\title{Actin driven morphogenesis in hydra}

\author{Sabyasachi Mukherjee}
 \altaffiliation{Physics Department, Indian Institute of Technology Bombay, Powai, Mumbai 400076, India.}
\author{Anirban Sain}%
 \email{asain@phy.iitb.ac.in}
\affiliation{Physics Department, Indian Institute of Technology Bombay, Powai, Mumbai 400076, India.}

\date{\today}

\begin{abstract}
Hydra, a centimeter long cylindrical-shaped freshwater organism, has emerged as an interesting model system for studying morphogenesis in animals.
Recently, fluorescent imaging of cytoskeletal actin
filaments on the outer surface of hydra has revealed nematic-type arrangement of actin filaments. {Several topological defects in the nematic field have also been detected.
In particular, aster-like +1 defects  appear at the curved head of hydra and at the tip of its tentacles, while -1/2 defects are seen at the base of the tentacles. However, functional role of these defects 
in tissue development is not clear.}
Motivated by these observations, we here model \textcolor{black}{hydra's
epthelial tissue as a visco-elastic membrane and the tentacles as growing
membrane tubes} driven by a nematic interaction among actin.  
We consider the epithelial layer of hydra as a fluid membrane 
and carry out a non-equilibrium simulation which also includes 
membrane growth and polymerization of actin. We show that specific 
kind of defect at the head does not play any positive role in 
emergence of the tentacles. The reorganization of actin at the 
base and the tip of growing tentacles are consistent with other
possible defect structures at the head as well.  
\textcolor{black}{While it is known that regions of tentacle growth are hot spots of chemical signaling, involving Wnt3/$\beta$-catenin pathway, we propose that active polymerization of actin bundles could also be an 
important player in the growth of tubular tentacles.  
In addition to polymerization}, \textcolor{black}{fluidity of our model membrane, capturing effective 
fluidity of the epithelial tissue,} turns out to be essential for 
enabling such growth.
\end{abstract}

\maketitle

\section*{Introduction}
Dynamic change of cell shape is integral part of biological 
morphogenesis leading to growth and differentiation of 
tissues into organs \cite{growthandform}. Cytoskeletal 
filaments, in particular actin fibers, assisted by actin binding proteins, form a dynamic network beneath the cell surface \cite{actinbook}. Movement of motor 
proteins (e.g., myosin) along actin fibers, powered by
energy released from ATP hydrolysis reaction, generates stresses 
in the actin network and causes change in cell shape. The
actin fibers  also get aligned in the process.
Following Ref\cite{natrphys} there has been a surge of interest 
\cite{headNoDefect2022,marchetti} in studying morphogenesis 
in hydra, a cylindrical shaped freshwater invertebrate, of radius
1 millimeter and length 10-30 millimeter. Using fluorescence microscopy Ref\cite{natrphys} visualized arrangement  of actin fibers on the outer epithelial layer  (ectoderm) of hydra. It has a cylindrical body shape with narrow tubular tentacles which grow from the upper end (the head) of its body. The actin fibers are aligned axially along the outer layer of the cylinder. Similar axial arrangement  was detected along the tentacles as well, see schematic Fig\ref{fig.summary}a. Further, Ref\cite{natrphys} interpreted the actin alignment as nematic rods arranged on a surface and identified topological defects (see Fig\ref{fig.summary}a), following the language of liquid crystals. They showed that the sum of all topological charges adds up to two, consistent with the Poincare index theorem (popularly known as "Hairy ball theorem")
for rod like objects, lying in the local tangent plane on a closed surface. In Ref\cite{natrphys} each tentacle was also shown to have a +1 defect at its tip which is charge neutralized (topologically) by two -1/2 defects near its base. \textcolor{black}{They had hypothesized \cite{natrphys} that the "unique mechanical environment" around a +1 defect may have "localized mechanical cues" for the head organizer. Based on this we  speculate that the resulting actin stress field around the aster like +1 defect may influence tentacle formation  since the head organizer dictates all downstream regenaration processes. We explore this possibility here.}
 \\


Earlier work had described emergence of new tentacles in hydra
as a result of  morphogen diffusion \cite{morphogen-diffuse1972} over its surface, generating Turing pattern \cite{Turing} like spots where from the tentacles emerge.
\textcolor{black}{While discussion of stresses, related to specific actin alignment and associated topological defects at the base and the tip of the tentacles, marks an interesting paradigm shift from morphogen centric chemical description to mechano-chemical description, chemical signaling
still remains an important factor for tentacle formation. }

\textcolor{black}{(a) High Wnt3/$beta$-catenin expression in the cell nuclei has been 
reported around the head, bud forming zone and in tentacles 
\cite{nakamura-betaCat-head-PNAS, natrbeta} of hydra. (b) Recent 
work \cite {sciAdv22-Zic4-betaCat} have shown that 
transcription factor Zic4, whose expression is regulated by 
Wnt3/$\beta$-catenin signaling and another transcription 
factor Sp5, are key to tentacle formation and its maintenance. 
(c) Over-expression \cite{overexpress} of $\beta$-catenin showed tentacle 
formation all over hydra’s body, including its main cylindrical 
trunk. In summary, the bud and tentacle forming regions of hydra, 
are hot spots of chemical activity which give rise to narrow tube 
like tentacles. Tentacle formation is accompanied by proliferation, differentiation, and migration of epithelial cells, and actin 
polyerization (see below), which is of interest to us here. 
Tentacles have myoneme cables and their major component is 
antiparallelly stacked actin filaments, which form via polymerization.} \\

\textcolor{black}{Actin-dynamics have been visualized in regenerating tissue. (a) Ref\cite{sea-anemone-actin-ten} 
studied actin dynamics during tentacle 
formation in sea-anemone, which belongs to the same cnidiria 
phylum as hydra, and concluded that actin polymerization 
are required for elongation of the body column and tentacles. 
(b) Continuous remodelling of cortical F-actin at the base of 
budding region has also been observed in hydra 
\cite{actin-dyn-Hobmayer}. Interestingly, dynamic actin-filled 
filopodia were also observed extending outward from the apical 
F-actin belt, and their possible role in re-organizing the 
contacts between neighbouring cells were highlighted. }\\

Theoretical modeling by various groups \cite{mahadevanPRL22, marchetti, juliahydra, geomi}, tried to understand the interplay among nematic defects, morphogen gradient, surface curvature and resulting stress, using active gel physics.   Ref\cite{mahadevanPRL22}
considered coupled dynamics of the nematic alignment field and the shape metric of an active curved surface and showed that positive (negative) curvature can occur near a +(-) defect. This is simultaneously accompanied
by area expansion (contraction) around a +(-) defect. Role of
activity has been shown to stabilize these defects. However,
how defects (with net zero charge), needed for the formation of a
tentacle, arise out of a defect free region near the head of 
the hydra is not clear. 
Ref\cite{marchetti} studied dynamics of actin alignment and defect unbinding on a predefined cylindrical surface with a 
hemispherical cap (the head of hydra). They assumed existence of a radial morphogen gradient, with concentration decreasing  away from the  head and showed that  an aster like +1 defect can be stabilized at the tip of the hemispherical head. They also assumed, phenomenologically, that the nematics couple with the morphogen gradient in a bilinear fashion. 
However, tentacle formation is a dynamic phenomenon that involves actin stress field, surface growth, and deformation; instead a predefined curved surface of fixed size and specified morphogen gradient was assumed in Ref\cite{marchetti}.  Simulations \cite{geomi,juliahydra} on the other hand focused on membrane protrusions promoted by the activity around motile +1/2 defects. The nematic arrangement around the protrusions in these cases are azimuthal in nature, as opposed to the axial arrangement seen on hydra.


\section{Model} 
\textcolor{black}{In our model active vertices are hotspots of enhanced chemical signalling resulting in physical shape change in the form of tubules, accompanied by specific axial arrangement of actin bundles. Although there is no direct
experimental evidence yet on how nuclear $\beta$-catenin may
induce actin polymerization, we speculate such a possibility. 
As mentioned in the introduction, Ref\cite{overexpress} have
shown that over-expression of $\beta$-catenin promotes tentacle
formation all over hydra's body and high proportion of 
nuclear $\beta$-catenin expression was found in the head and tentacles \cite{nakamura-betaCat-head-PNAS, natrbeta} of hydra.} 
We propose a model incorporating actin bundle formation 
as an active phenomenon. However unlike in Ref\cite{marchetti} where a predefined morphogen gradient ($\beta$-catenin in this case) on a predefined curved surface decides nematic (actin) alignment, in our model the morphogen density merely decides the site of actin growth and bundling. This growth subsequently decides the tubular shape of the tentacles. Our model 
introduces the active role played by the growing actin bundle in shaping the surface rather than adjust their orientation on an already curved surface. It is important here to recognize that the long actin cables in the ectoderm of hydra are not like
the constantly poly/depoly-merizing actin network found in cell cortex. The actin cables in ectoderm are clearly correlated over long length scales (see Ref\cite{natrphys}) and are likely to be more rigid compared to that in cell cortex. So far as the alignment is concerned the actins in ectoderm can still be treated as a nematic field, however due to their stiffness and large persistence 
length they are likely to play a much stronger role in the shape determination in hydra. For example, growth of actin
bundles (via polymerization) and subsequent growth of epithelial tissue could be the main driving force here, which is not taken into account in active gel physics.\\
\begin{figure*}
    \centering
    \includegraphics[width=0.8\linewidth]{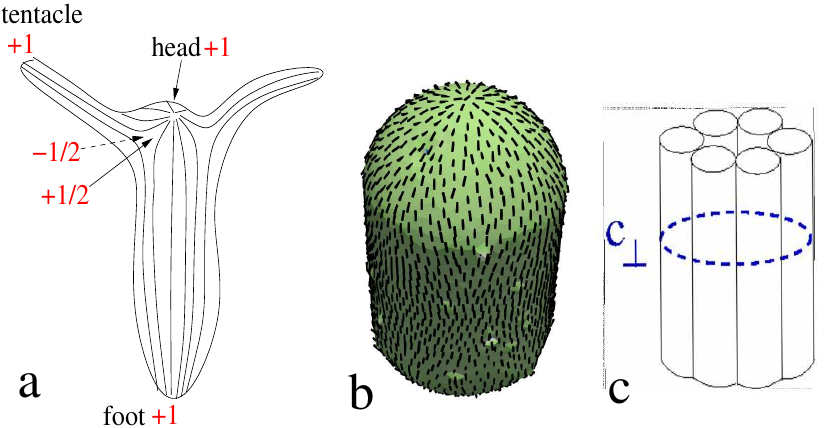}
    \caption{ 
a) Schematic picture of aligned actin fibers in the surface layer  of a hydra \cite{natrphys}. The alignment, when interpreted as a nematic orientation field, shows topological defects which adds
up to total charge 2. Each tentacle is charge neutral, with one
+1 defect at its tip and two -1/2 defects at its base. The 
dotted arrow indicates that the defect is on the back side.  b) Initial vesicle shape, used for our simulation, with two +1 defects held at the two poles. c) The $c_{\perp}$ term of our model promotes axial arrangement of nematics around a narrow tube of radius $c_\perp^{-1}$}.
    \label{fig.summary}
\end{figure*}

Following experimental observation \cite{natrphys}, 
we construct a minimal body plan for hydra, which we use to start the simulation. Ref\cite{natrphys} showed that the actin fibers are 
arranged parallel to the axis along its long cylindrical trunk and they converge to two aster like +1 defects at the two caps
of the cylinder. The upper cap is nearly hemispherical 
and hosts the tentacles while the lower cap at the 
foot is flat and these two +1 defects are held fixed
during morphogenesis of the hydra. We mimic this geometry by considering a short, cylinder shaped vesicle (the body), capped by a hemispherical cap at the top and a nearly flat bottom (see Fig\ref{fig.summary}b). The nematic field is held parallel to the axis on the cylindrical part. Except at the upper hemispherical cap, a high bending modulus ($\kappa$) is assigned to the rest of the vesicle
 so that it resist deformation leaving only
the upper hemispherical part to deform. \textcolor{black}{Recent experiment \cite{ECM-diff-siffness} shows inhomogeneous stiffness pattern along hydra's body axis. Its upper part (towards the head) is softer while the lower part (towards the foot) is stiffer. }
\textcolor{black}{We assume, even for a regenarating hydra, this feature is qualitatively the same, resulting in a relatively softer extra cellular matrix (ECM) near the head which host the tentacles.} 
In some of our simulations we additionally hold an aster like +1 defect at the head. This is done by fixing the nematic orientations on two ring of vertices around the top most vertex of the hemispherical cap. This amounts to fixing both the positions and the nematic orientations at the topmost vertex and its nearest and next nearest neighbor vertices. Rest of the vertices in the cap are allowed to change their spatial position and nematic orientation during the simulation. We also maintain 
high bending rigidity ($\kappa$) for the central and the two 
rings of vertices constituting the head, in order to keep
this small region undeformed. \\

In our model the membrane surface is represented by a 
discrete, triangulated, closed network \cite{SunilPRE} of vertices.  Each vertex (numbered by $i$) has a nematic director $\hat n_i$ lying on the local tangent plane of the membrane  \cite{SunilPRE} and is free to rotate in this plane. The director influences the local curvature at the vertex 
by imposing differential intrinsic curvature along its 
own orientation and perpendicular to it, while the membrane opposes
any bending due to curvature. 
In our Monte-carlo simulation 
there are three equilibrium Monte-Carlo moves: a) vertex shifts, b) rotation of the nematic in the local 
tangent plane of a vertex, and c) link flip. The first move (a) updates the vertex position keeping the triangulation fixed and thus is responsible for shape change of the membrane.  (b) is responsible for appropriate nematic alignment on the curved surface that lowers the nematic part of the energy. 
In the link flip move, a link connecting two vertices 
is picked at random and then erased, and subsequently 
a new link is established between a pair of vertices 
which were on the opposite sides of the erased link.
This move changes the nearest neighbors of any given 
vertex and cause diffusion of the vertex in the network
and renders fluidity to the membrane. Details of all 
these Monte-Carlo moves leading to vesicle shape 
deformation and corresponding nematic arrangement, 
while 
maintaining membrane fluidity, are available in Ref\cite {SunilPRE, Sunilbiopj,tubepre}.  \textcolor{black} {Note that effective membrane fluidity can arise on long time scales 
due to cell sorting and spreading \cite{sorting-fluidization}, T1-transitions (intercalation) \cite{T1-fluidization}
or cell division and apostosis \cite{julicher-fluidization}.
Tentacle growth can involve any or all of these processes.}  \\

In addition to energy considerations we also incorporate activity
into the model following a recipe, 
partly similar to that proposed in Ref\cite{Madanrao}. 
In this recipe, in addition to moving the positions of the vertices (using MC moves that are aimed at energy minimization) we introduce activity dependent transition rates in the model that do not obey detailed balance.  This is analogous to nonequilibrium Langevin equations where the dynamics depend on both thermodynamic forces obtained from underlying energy function of the system and active noise that do not obey detailed balance.\\

To model the thermodynamic forces we use the following modified Helfrich-Canham Hamiltonian \cite{SunilPRE, Sunilbiopj,tubepre} 
which includes a nominal isotropic bending rigidity ($\kappa$), 
the anisotropic membrane bending rigidities ($\kappa_{\parallel}$ and $\kappa_{\perp}$) induced by the local nematic direction and the alignment interaction ($\epsilon_{LL}$) among the nematics. 
\begin{align}
 H &= \sum_i \Bigg[ \frac{\kappa}{2}(2H_i)^2
     + \frac{\kappa_{\parallel}}{2}(H_{\parallel,i}-c_{\parallel})^2
     + \frac{\kappa_{\perp}}{2}(H_{\perp,i}-c_{\perp})^2 \Bigg] \nonumber \\
   &\quad - \epsilon_{ll}\sum_{i>j} \big[ (\hat{n_i}\cdot \hat{n_j})^2 \big]
\label{eq.H}
\end{align}

In this discretized energy function $H_i$ is the mean curvature 
at the $i-$th vertex. $H_i=(c_{1,i}+c_{2,i})/2$, where $c_{1,i}$ 
and $c_{2,i}$ are the local principal curvatures on the membrane surface along orthogonal tangent vectors $\hat t_{1,i}$ and $\hat t_{2,i}$. Further,  $\kappa_{\parallel}$ and $\kappa_{\perp}$ 
are the membrane bending rigidities parallel and perpendicular 
to the nematic axis respectively.
$H_{\parallel,i}$ and $H_{\perp,i}$ are the local membrane curvatures parallel and perpendicular to the local nematic  axes, $\hat n_i$ and $\hat n_{\perp,i}$ respectively, while $c_{\parallel}$ and $c_{\perp}$ are the corresponding intrinsic curvatures induced by the nematic director. The last term of the Hamiltonian is the standard Lebwohl-Lasher \cite{LL} term (in one constant approximation), promoting alignment between the nematics at all neighboring vertex pairs $(i,j)$. Since actin does not have any intrinsic curvature we will set $c_{\parallel}=0$. 
But we set $c_{\perp}>0$ to promote development of narrow membrane 
tubules induced by growing actin bundles. Note that the term $c_{\perp}$ is the preferred curvature of the membrane perpendicular to its nematic axis. Such a curvature can be imposed on the membrane
by a growing bundle of actin which pushes out the membrane from 
inside of the vesicle into a tubular shape. Alternatively, actin can align axially on the outer surface of a membrane tube, a) due to 
their mutual nematic interaction, and b) if there are lateral
attraction between the actin filaments. Fig\ref {fig.summary}c shows 
how the bundling effect takes place due to the $c_{\perp}$ term.
This strategy was originally developed and explored in Ref\cite{tubepre}. 
In the present context both the factors (a,b) are likely to be responsible for actin organization in the tube. 
\textcolor{black}{As mentioned earlier \cite{nakamura-betaCat-head-PNAS, natrbeta, betacatenin2019}, Wnt3/$\beta$-catenin signaling is high near the head region.}

The local membrane curvature along $\hat n_i$ and $\hat n_{\perp,i}$ are given by $H_{\parallel,i}=c_{1,i} \cos^2 \phi_i + c_{2,i} \sin^2 \phi_i$ and $H_{\perp,i}=c_{1,i} \sin^2 \phi_i + c_{2,i} \cos^2 \phi_i$, where $\phi_i$ denotes the angle between the director $\hat n_i$ and principal direction  $\hat t_{1,i}$. Note that in order for the Hamiltonian to have rotational symmetry (frame invariance), it must be function of the invariants of the curvature tensor $C_{ij}=\partial_i n_j$ and the vector $n_i$. One can
show that all the anisotropic terms in the Hamiltonian 
can be expressed in terms of the invariants. For example, $H_{\parallel}=Tr(C\hat n\hat n)= C_{ij} n_j n_i$. \\

We introduce two type of vertices, active and
inactive, denoted by '+' and '-' markers, respectively. The active vertices have $c_\perp\neq 0$ on them, while the inactive vertices have  $c_\perp = 0$. A vertex 
can switch between these two states. We introduce on-rate 
$P(-+)$ and and off-rate $P(+-)$, for transitions from $- \rightarrow +$ and $+ \rightarrow -$, respectively. By active state, here we mean vertices that are capable of promoting or participating in bundle formation via polymerization/ myosin mediated sliding. \textcolor{black}{We assume active vertices are the ones where Wnt3/$\beta$-catenin signaling is high}.  Here  $N$ is the total number of vertices, while $N^+$ and $N^-$ are the active and inactive vertices, respectively, with $N_+ + N_-=N$. We use the  following conversion rates, 
\begin{eqnarray}
P (-+)&=&\Big( \frac{N^+}{N} \frac{1}{1 + e^{\mu(A-A_0)}}\Big)\frac{n^+}{n} \\
P (+-)&=& \Big(\frac{N^-}{N} \frac{1}{1 + e^{-\mu(A-A_0)}}\Big )\frac {n^-}{n} 
\label{eq.p+-}
\end{eqnarray}

\begin{figure}
    \centering
    \includegraphics[width=0.5\linewidth]{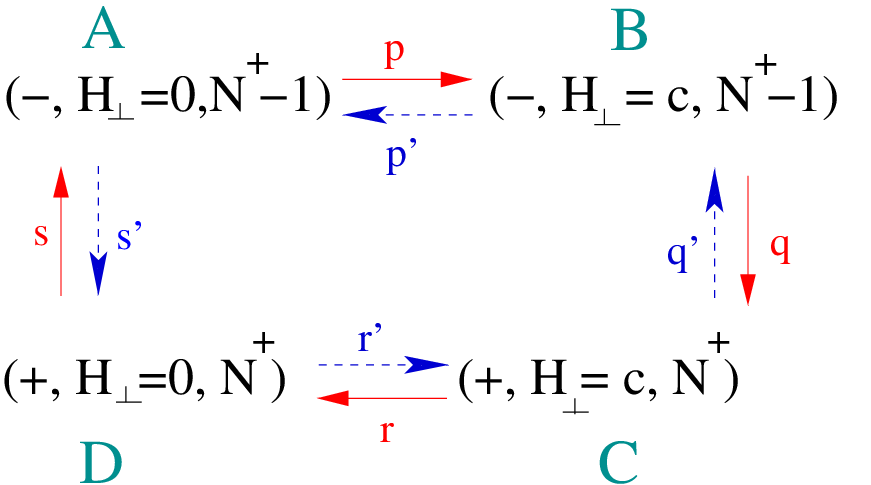}
    \caption{The loop consisting of 4 states A,B,C and D does not satisfy Kolmogorov criteria for detail balance. Here $pqrs\neq p'q'r's'$. For the transition rates $p,q,r,s$ and $p',q',r,s'$, see text. }
    \label{fig-kolmogorov}
\end{figure}

These rates are product of two factors. The first factor in the parentheses depends on the instantaneous global quantity $N^+$.
It drives the population difference $A=N^+-N^-$ towards a desired value $A_0=N_0^+-N_0^-$, where $N_0^+$ and  $N_0^-=N- N^+_0$ are the desired values of $(N^+, N^-)$. With this factor alone the system
can reach a steady state $N^-P(-+)=N^+P(+-)$ for $A=A_0$.
 The second factor  $n^+/n$ (or $n^-/n$) depends on the local environment of a vertex.  Here $n$ is the number of nearest neighbors of a vertex (which ranges between 5 to 7) and $n^+$ (and $n^-$) is the number of '+' (and '-') vertices among $n$, with $n^+ + n^-=n$. 
The term $n^+/n$ favors a vertex to turn active if majority of its nearest neighbors are active (i.e., if a bundle is forming in its vicinity) and similarly turn inactive if majority of its neighbors are inactive. Since active vertices promote bundling, these rules model the actin bundling as an autocatalytic event \cite{autocatalytic}. It promotes formation of active patches which subsequently give rise to tubes. Due to the second factor it is not obvious whether the system can reach a steady state; at least $N^-P(-+)=N^+P(+-)$ does not hold true. The transition rates here do not depend on the local energy but break pairwise detail balance between micro-states, as described by $(+/-,H_\perp,N^+)$ in Fig.\ref{fig-kolmogorov}. But detail balance can still be recovered for Kolmogorov loops, retaining the system at equilibrium.\\

Now we show that the simulation scheme that we constructed above, drives the system out of equilibrium. Towards this, we 
form a closed loop of 4 states which are connected by successive transitions that follow either Monte-Carlo moves (dictated by energy) or non-equilibrium transition ($-\rightarrow +$ or reverse). All the 4 states correspond to  
the changes that occur at a single vertex, indexed by its activity (+ or -), its $H_\perp$ value and the global state 
$N^+$.  In Fig.\ref{fig-kolmogorov} the 4 states are denoted by A,B,C and D and the loop is traversed once along the solid arrows (clockwise) and another time along the dashed arrows (counter-clockwise). In A and B the vertex is inactive (-) 
and hence its contribution to the curvature energy is 
$\frac{\kappa_{\perp}}{2}H_{\perp}^2$, where as in C,D its
contribution to the curvature energy is 
$\frac{\kappa_{\perp}}{2}(H_{\perp}-c_{\perp})^2$. We consider B and C to have a positive curvature $c_\perp > c>0$ and A and D to have none. Therefore,
for transition $A\rightarrow B$ curvature energy increases :     $0\rightarrow \frac{\kappa_{\perp}}{2}c^2$. Similarly, also for transition $C\rightarrow D$ the curvature energy 
increases : $\frac{\kappa_{\perp}}{2}(c-c_\perp)^2 \rightarrow \frac{\kappa_{\perp}}{2} c_\perp^2$. Thus, 
\begin{eqnarray}
p&=&e^{-\Delta E/k_BT},\; p'=1\;\mbox{where}\; \Delta E= E_B-E_A>0  \nonumber\\
r&=& e^{-\Delta E/k_BT},\; r'=1,\;\mbox{where}\; 
\Delta E=E_D-E_C >0
\label{eq-eql}
\end{eqnarray}
Note that, energy also changes during the transitions $B\rightarrow C$ and $D\rightarrow A$ but these transitions 
are not governed by energy change but by the non-equilibrium 
rates $P(-\rightarrow +)$ and $P(+\rightarrow -)$, respectively.
In Fig.\ref{fig-kolmogorov} we specifically have,
\begin{eqnarray}
q&=&s'= P(-+)=\Big( \frac{N^+-1}{N} \frac{1}{1 + e^{\mu(A-A_0)}}\Big)\frac{n^+}{n} \nonumber\\
s&=&q'= P(+-)=\Big(\frac{N-N^+}{N} \frac{1}{1 + e^{-\mu(A-A_0)}}\Big )\frac {n^-}{n}
\label{eq-noneql}
\end{eqnarray}
During these transitions, the local environment $n^+, n^-$
of the particular vertex are not changed. It turns out that the product $pqrs$ (for clockwise rotation) differ from $p'q'r's'$ (for anticlockwise rotation). Here $qs=q's'$, but $pr\ne p'r'$.
Note, that the transitions between A-D or B-C do not follow pairwise detail balance although their contributions to the Kolmogorov loop turns out to be same
(i.e., $qs=q's'$). But even the transitions between A-B and C-D 
do not preserve pairwise detail balance here because the populations of the states A and B, for example, are not proportional to $e^{-E_A/kT}$ and $e^{-E_B/kT}$. Even when $c_\perp >0>c$, as opposed to the case considered above, one can show that $pqrs\ne p'q'r's'$. In
this case $E_B>E_A$, but $E_D< E_C$, since $\frac{\kappa_{\perp}}{2}(|c|+c_\perp)^2 >\frac{\kappa_{\perp}}{2} c_\perp^2$.\\

The simulation was started with $N^+_0$ active vertices randomly distributed on the upper hemispherical cap (excluding the head) consisting of total $N$ vertices. The initial nematic orientations were chosen randomly except that an aster like +1 defect was imposed on the head. The head was defined as the topmost vertex of the initial hemisphere and its nearest and next nearest neighbor vertices. The nematic director at the topmost vertex (center) was assigned an arbitrary orientation since it is a singularity 
where nematic orientation is undefined. The directors on the two surrounding ring of vertices were chosen to point towards the central vertex yet lying in the local tangent plane. For this, the link connecting the neighboring vertex to the central vertex was projected onto the local tangent plane of the neighboring vertex. The director was placed along this projected line and was held fixed during the simulation.  
While this recipe was used for fixing the aster like +1 defect, for the ring like +1 defect the directors were made to point towards their neighbor along the same ring. Neither MC moves
(which alter positions of vertices) nor bond flip moves   (responsible for implementing membrane fluidity) were permitted for these set of vertices in order to keep their connectivity and the +1 defect intact. \\

\begin{figure}
    \centering
    \includegraphics[width=0.5\linewidth]{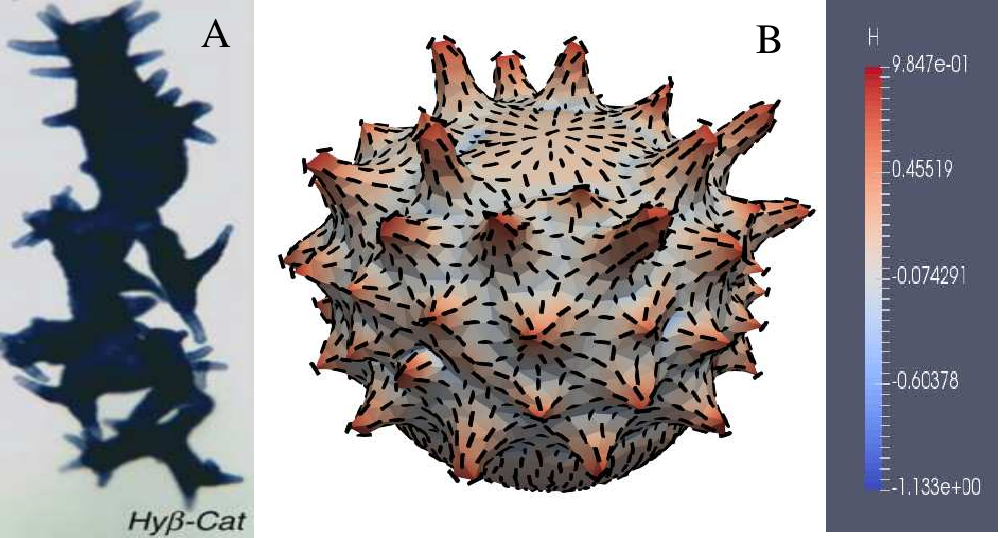}
    \caption{Over expression of $\beta-$catenin promotes
    tentacle formation even on the cylindrical portion
    of hydra : A) Hy$\beta$-cat in $\beta$-cat Tg (Transgenic Hydra with fluorescently labeled $\beta$-catenin), taken from Ref\cite{overexpress}. B) Growth of tubes when all vertices are active, in our simulation. The scale bar shows local curvature values. }
    \label{fig-overexpress}
\end{figure}

The active-inactive transitions on vertices were attempted sequentially with MC moves which alters the spatial position of the vertices. After every MC sweep 
a fixed percentage of the vertices were randomly chosen and their activity states were changed with the probabilities $P(+-)$ or $P(-+)$. 
The active vertices self organized into clusters due to the local component of the transition rates.  Each such cluster of vertices deformed into tubes. Quite trivially, when nonzero $c_{\perp}$ was switched on for all vertices, including the ones on the cylindrical trunk, tubes grew all over the body (see Fig.\ref{fig-overexpress}B), similar to that seen in experiments (Fig.\ref{fig-overexpress}A) 
where $\beta$-catenin was over-expressed.\\


\section{Results and Discussion}
\begin{figure*}
    \centering
    \includegraphics[width=0.8\linewidth]{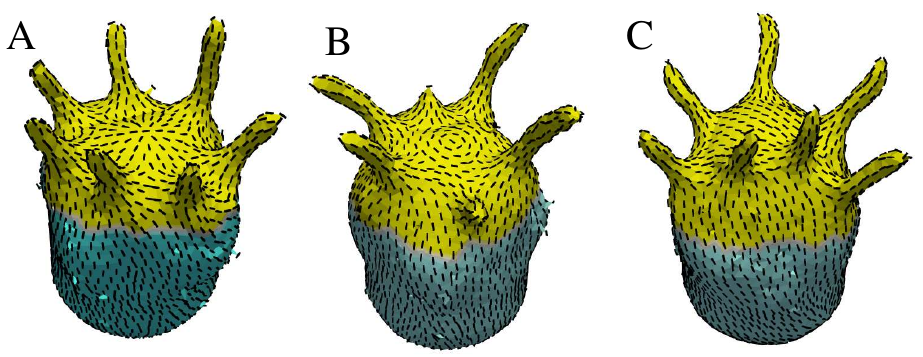}
    \caption{Tentacles resulting from non-equilibrium simulations. (A) Aster like +1 defect, and (B) ring 
    like + 1 defect, held at the head. In (C) no defects are imposed at the head. The active and inactive vertices 
    are shown in different colors, yellow and blue, respectively.
    Parameters: $N_0^+/N=0.9, c_{\perp}= 1.0 ,\kappa = 10 , k_{\perp}= 30, k_{\parallel}= 20, \epsilon_{ll}= 6 $    }
    \label{fig-noneqlb}
\end{figure*}

It is generally expected that topological constraints
like the Poincare-Hopf theorem (popularly known as the "Hairy ball theorem" impose long-range correlation in the nematic alignment by constraining the total defect charge on a closed surface. \textcolor{black}{Ref\cite{natrphys, headNoDefect2022} had highlighted how the +1 defect may create
localized mechanical cues for the head organizer. Since the head organizer dictates all downstream regeneration processes, question
arises whether the unique location of the +1 defect dictates 
positioning of the tentacles as well. }
To check this we ran our simulations for  three different cases, a) with an aster like +1 defect, b) with a ring like +1 defect and c) with no defect, held at the head, see Fig.\ref{fig-noneqlb}.
We conclude, that the formation of tubular tentacles around the head region is insensitive to the nature of +1 defect, i.e., whether it is aster like 
(as in our Fig.\ref{fig-noneqlb}A) or ring like (as in Fig.\ref{fig-noneqlb}B). Furthermore, tubes can 
form even in the absence of any defect at the head as 
shown in Fig.\ref{fig-noneqlb}C. \\

The  tubes are deformations accompanied by a nematic field with zero net topological charge; the top of the tube
has a aster like +1 charge while there are two -1/2 
charges located near the base of the tube. These three charges are created simultaneously (defect splitting) in a small area which
previously had a uniform nematic alignment. The +1 defect separates out from the other two by elongation
of the tube which requires polymerization of actin and growth of the  membrane. In terms of the actin alignment we hypothesize that this defect splitting may be caused by a split in a growing actin fibers or fiber bundles and a local rearrangement of fibers around the split. Since the local net charge is still zero such simultaneous generation of -1/2,+1,-1/2 triplet
will not have any long distance effect over the rest of
the curved surface. \textcolor{black}{The tentacle evolution is captured 
in two  movies (Mv-1 and  Mv-2 in supplementary information) exhibiting, (1) self-organization of the
active vertices, and 2) defect nucleation and tentacle growth. }\\

\begin{figure*}
    \centering
    \includegraphics[width=1.0\linewidth]{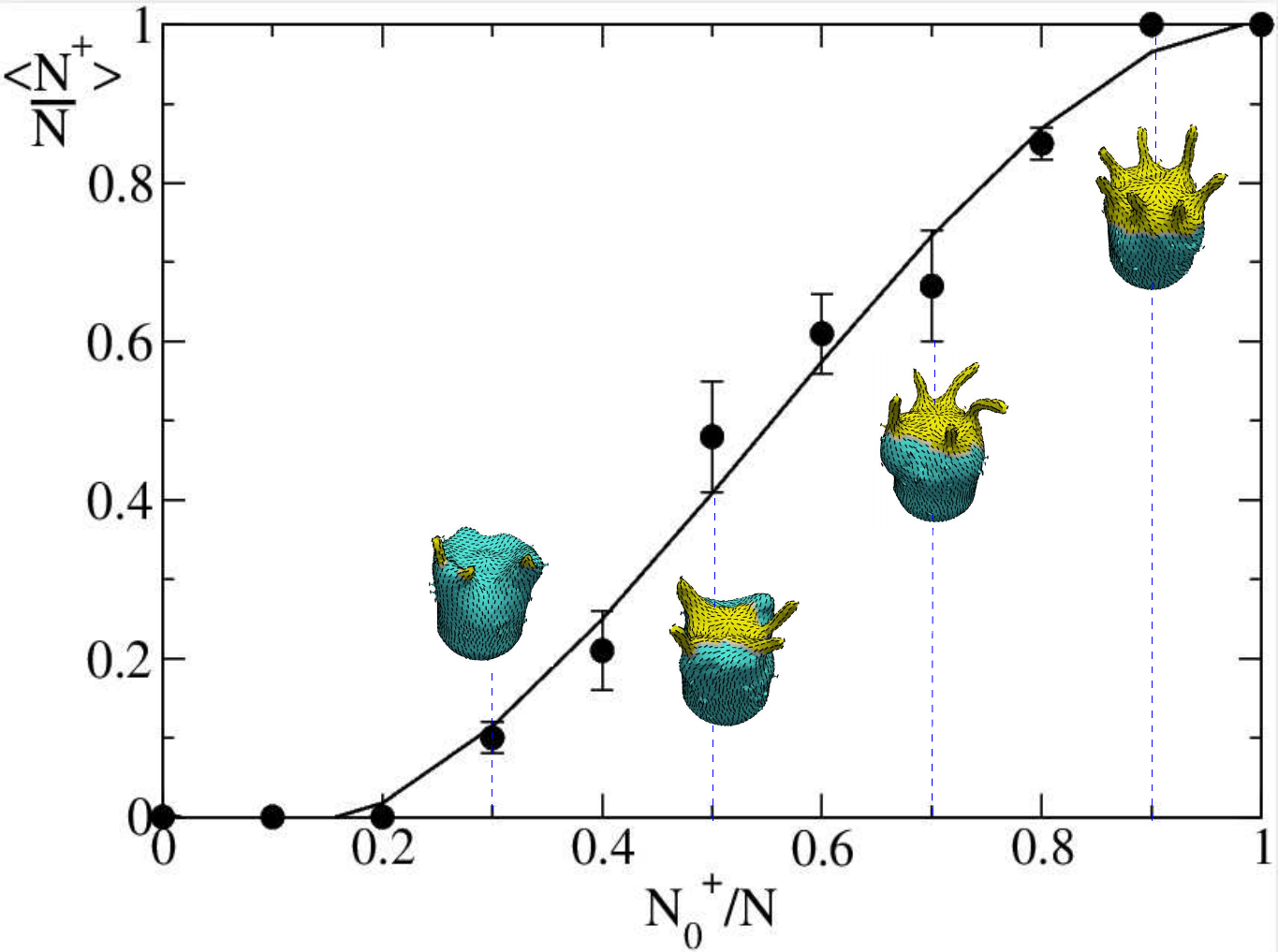}
    \caption{Average fraction of active vertices in the non-equilibrium steady state, as a function of initial fraction of active vertices. The sigmoidal dependence implies that a threshold fraction of active vertices is needed for tentacle formation. Error bars are computed from 5 simulation runs per data point.}
    \label{fig.calibrate}
\end{figure*}

The percentage of active vertices ($N^+$), in our simulation, can be qualitatively interpreted as the density of active sites where Wnt3/$\beta$ catenin signalling is high. 
The fraction of active sites 
$N^+/N$ decides the number of tubes. This is because tube formation requires self assembly of active vertices
which form patches at different locations. Each patch 
potentially generates a tube.
Although we start the simulation with $N^+=N_0^+$, in the
nonequilibrium steady state $N^+$ differs from $N_0^+$. 
During the simulation the population difference {\bf $A=N^+-N^-$} fluctuates. The global parts of the rates $P(-+)$ and $P(+-)$
are designed to maintain $\langle A\rangle= A_0$ (equivalent to 
$\langle N^+\rangle= N_0^+$. But The local components of the rates ($n^+/n$ and $n^-/n$) breaks this statistical conservation,
i.e., $\langle N^+\rangle\neq N^+_0$. In Fig.\ref{fig.calibrate} we plot the emergent $\langle N^+\rangle/N$  as a function of the input parameter $N^+_0/N$. The corresponding 
steady state configurations are also shown along with some of the data  points. The ratio $\langle N^+\rangle/N$ increases monotonically with $N^+_0/N$ following a sigmoidal curve. The solid line in the figure is not
a best fit but a guide to eye to show the nonlinearity. 
It is clear from Fig.\ref{fig.calibrate} that tubes cannot form below a  threshold $N^+_0/N\approx 0.5$. This can be interpreted as actin bundling requiring a \textcolor{black}{critical concentration of active sites.}  For example, when $N^+_0/N < 0.5$, the resulting $N^+$ is suppressed because of the the local dominance of inactive vertices, which bias the conversion rate in our local moves.\\



What controls the number and diameter of the tubes ? We discussed above how $N^+_0$ decides resulting $N^+$ and influence both number and length of tubes. But  $c_{\perp}$ and $\epsilon_{LL}$ also have strong influence. $c_{\perp}$ is directly proportional to the inverse tube radius. Higher 
the $\epsilon_{ll}$ lesser is the number of tubes because tubes
have higher nematic elastic energy cost. Membrane elastic energy cost 
($\kappa$) is also higher for tubes which has high mean curvature. 
Some of these effects were demonstrated in our earlier work \cite{tubepre}.  \\
One essential ingredient of our simulation turns out to be the 
link flip moves which renders fluidity to the membrane. Link flip 
moves allow a pair of nearest neighbor vertices to delink and become
next nearest neighbor on the network, and simultaneously a pair of 
vertices which were next nearest neighbors become nearest neighbors.
These adjustment of links has two effects on the energetics. First, 
it relaxes elastic stresses, due to excessive elongation or shortening 
of links between neighboring vertices. Second, vertices can diffuse 
through the network mimicking flow on longer time-scale. Since the
elastic interactions among the nematics in our model are implemented
via nearest neighbor interactions (as is standard in the discrete LL-model \cite{LL}), nematic defects can be easily created. 
Once defects are created tubes can grow by taking advantage of the 
$c_{\perp}$ term in the Hamiltonian which lowers energy costs by
arranging the nematics parallel to the tube axis, provided the 
adjoining vertices are in active state. This axial arrangement 
of nematics can be interpreted as active polymerization. \\

Further,
the growth of tube requires new vertices to be lifted off from
the base of the tube which in turn requires supply of new vertices.
In our simulation new vertices are continuously drawn from the cylindrical zone to the head zone, as it becomes energetically favorable for a vertex to move onto the tube than staying back
in the cylindrical zone. This mimics growth in our model. Most
importantly, we noted that in the absence of link flip moves 
the tubes do not grow. Thus, fluidity of the membrane, in our 
model is linked with defect unbinding, active polymerization and
growth of the tissue. This indicates that the elasticity of the epithelial tissue is not sufficient for tentacle growth,
\textcolor{black}{but tissue fluidity is also necessary. }  \\

Note, that activity is brought into our model through the active vertices which can promote actin polymerization or bundling. 
active stresses have nonequilibrium origin and are non-isotropic, biased by the local orientation of the nematic 
director, which is often actin. In contrast, in our model the 
source of stresses that drive tubular growth comes from the 
Hamiltonian, however the switching on/off rates of these active 
sites (and hence polymerization) and their self-assembly are  non-equilibrium in nature.\\

\textcolor{black}{The $c_{\perp}$ term in our hamiltonian allows for
axial nematic (actin) alignment (as opposed to azimuthal alignment) on cylindrical tubes of preferred radius. However, this term does not a priori tell us, a) how many tubes (equivalently, inter tube distance) and of what length will emerge (the emergent number here nearly agrees with reality). b) How nematic alignment at the base of the tube interact with the nematic alignment influenced by the defect imposed at the head. This brings out one of the main conclusions of the study that the type of defect at the head (and even no defect) does not interfere in tentacle formation. c) There is possibly a threshold in the required fraction of active sites for tentacles to emerge, and d) the requirement for membrane (tissue) fluidity.}

In summary, \textcolor{black}{we have proposed chemical signaling  assisted actin bundle formation, which could be an active process, and which combines with proliferation, differentiation and migration of epithelial cells to form tentacles. As we showed,
the positioning of the  +1 defect at the center of the head merely acts like a place holder where tubes do not form}.
We implemented this by assigning a high value of $\kappa$ and setting $c_{\perp}=0$ on two rings of vertices around the
central vertex at the head. We showed that the defect is not the driving force for tube formation. 
The tube formation requires simultaneous generation of
a defect triplet locally with charges $-1/2,+1,-1/2$,
and subsequent separation of these defects. Since
their total charge is zero this defect unbinding does not 
disturb global constraints on the total charge
as dictated by the hairy ball theorem. Once the defects 
are formed Ref\cite{mahadevanPRL22} has shown how, via
nematic-metric interaction, positive curvature and 
growth takes place at positive defects. Our simulation
gives a pathway for nucleation of these defects (via 
the inherent stochasticity involved in Mont-Carlo 
simulation). The defects are further stabilized, 
leading to growth of tubes by the nematic bundle 
promoting term  ($c{\perp}$) in the energy function.
In reality, the actin filaments in hydra form long
contiguous fibers as visualized in Ref\cite{natrphys}.
This is unlike that in cell cortex where meshwork 
of short filaments forms a loosely cross-linked 
network. It is possible  that 
the contiguous actin fibers may split and reorganize 
to form defects and \textcolor{black}{then respond to cues 
from local chemical signaling to polymerize and remodel 
along side  migrating epithelial cells to form tentacles.
In this scenario the positive defects form first which 
subsequently leads to convex curvature and growth \cite{mahadevanPRL22}. But it could also be possible
that a spurt of signaling activity may initiate out of 
plane actin (and epithelium) growth. }
Such an actin bundle may reorganize on the surface of the tentacle conforming to -1/2,+1,-1/2 defect triplet. 
To distinguish between these two scenarios we need experiments where growth and defect formation can be decoupled, for example by suppressing actin bundling proteins. 
 

\section{Author Contribution}
SM performed the research, analyzed the data, and also participated in the design of the research. AS designed and supervised research and participated in data analysis. SM and AS wrote the manuscript together.

\section{Acknowledgements}
A.S. acknowledges the Science and Engineering Research Board (SERB), India (Project Nos. CRG/2019/005944) for financial support.
\bibliography{refs}
\end{document}